\documentclass[letterpaper, 12pt]{article}
\usepackage{booktabs, array, caption, natbib}
\usepackage{amsfonts,mathrsfs}
\usepackage{booktabs}
\usepackage{longtable}
\usepackage{rotating}
\usepackage{float}
\usepackage{caption}
\usepackage{color}
\usepackage{physics}
\usepackage{setspace}
\usepackage[margin = 1in]{geometry}
\RequirePackage[colorlinks,citecolor=blue,urlcolor=blue,linkcolor=blue]
{hyperref}

\begin{document}
\newcommand{\U}{{\boldsymbol{U}}}
\newcommand{\A}{{\boldsymbol{A}}}
\newcommand{\G}{{\boldsymbol{G}}}
\newcommand{\V}{{\boldsymbol{V}}}
\newcommand{\R}{{\boldsymbol{R}}}
\newcommand{\Q}{{\boldsymbol{Q}}}
\newcommand{\bH}{{\boldsymbol{H}}}
\newcommand{\br}{{\boldsymbol{r}}}
\newcommand{\s}{{\boldsymbol{s}}}
\newcommand{\bxi}{{\boldsymbol{\xi}}}
\newcommand{\T}{{\bf{{T}}}}
\newcommand{\1}{{\bf{{1}}}}
\newcommand{\X}{{\boldsymbol{X}}}
\newcommand{\bbeta}{{\boldsymbol{\beta}}}
\newcommand{\btheta}{{\boldsymbol{\theta}}}
\newcommand{\0}{{\bf{{0}}}}
\newcommand{\I}{{\boldsymbol{\mathcal{I}}}}
\newtheorem{thm}{Theorem}[section]
\newcommand{\proglang}[1]{\textsf{#1}}
\newcommand{\pkg}[1]{{\normalfont\fontseries{b}\selectfont #1}}
\newcommand{\blue}[1]{{\textcolor{blue}{#1}}}
\newcommand{\red}[1]{{\textcolor{red}{#1}}}
\renewcommand{\bar}{\overline}
\renewcommand{\hat}{\widehat}
\renewcommand{\tilde}{\widetilde}

\title{An Online Updating Approach for Testing the Proportional
Hazards Assumption with Streams of Survival Data}

\author{Yishu Xue\thanks{Department of Statistics,
University of Connecticut, Storrs, CT 06269, USA} ~~
 HaiYing Wang\footnotemark[1] ~~ Jun Yan\footnotemark[1] ~~ 
Elizabeth D. Schifano\footnotemark[1]~\thanks{(to whom correspondence
should be addressed) Email: elizabeth.schifano@uconn.edu}}

\maketitle

%  put the summary for your paper here

\begin{abstract}
The Cox model, which remains as the first choice in analyzing time-to-event
data even for large datasets, relies on the proportional hazards (PH)
assumption.
When survival data arrive sequentially in chunks, a fast and minimally storage
intensive
approach to test the PH assumption is desirable.
%When the data volume exceeds the computer memory, the standard statistics for
%testing the PH assumption can no longer be easily calculated.
We propose an online updating approach that
updates the standard test statistic as each new block of data becomes
available, and greatly lightens the computational burden.
Under the null hypothesis of PH, the proposed
statistic is shown to have the same asymptotic distribution as the standard
version computed on the entire data stream with the data blocks
pooled into one dataset.
In simulation studies, the test and its variant based on most recent data
blocks maintain their sizes when the PH assumption holds and
have substantial power to detect different violations of the PH
assumption. 
We also show in simulation that our approach can be used successfully with 
``big data'' that exceed a single computer's computational resources.
The approach is illustrated with the survival analysis of
patients with lymphoma cancer from the Surveillance, Epidemiology, and End
Results Program. The proposed test
promptly identified deviation from the PH assumption that
was not captured by the test based on the entire data.
\bigskip

\end{abstract}

\section{Introduction}
\label{s:intro}

Recent advances in information technology have made available data that
arrive in high velocity everyday. Online methods, such as the online updating
estimation and inference presented in \cite{schifano2016}, are appealing
as storage of historical data is not required which yields great savings in
computing resources. 
Survival data, or time-to-event data, may also arrive sequentially, and the 
desire for online updated inferences in the survival setting is not
 uncommon.  For example, flight information, such as delay time until
 take-off or cancellation, is available for more than 114,000 commercial
flights
 scheduled daily around the world~\citep{aviation2018};
 real estate information, such as time on market until sold, is updated
 continuously for the over 6 million homes
 in the real-estate market~\citep{estate2018}. As such events occur everyday
at high frequency, observations also accumulate quickly.

%With saving
%summarized information from subsets of data and aggregating them to produce
%the final result, the estimation and inference can be updated promptly as
%new data arrive. Survival data, with its unique time-to-event structure,
%has been of great research interest. It exists not only in the public
%health research field, but also in many others. For example, flight
%information, such as delay time until take-off or cancellation, is
%available for more than 114,000 commercial flights scheduled daily around
%the world \citep{aviation2018}; real estate information, such as time
%until sold, is updated continuously for the over 6 million homes in
%the real-estate market \citep{estate2018} . 

The Cox model \citep{cox1972} is the most commonly used tool in analyzing
survival data. % and remains so even for massive data
%\citep[e.g.,][]{Mittal2014}.
A crucial step in fitting the popular Cox model is to check the proportional
hazards (PH) assumption \citep[e.g.,][]{xue2017}.
The standard approach, if new data becomes available along a stream, would be
to pool all historical data together, fit a new Cox model, and use standard
methods such as the test of \cite{grambsch1994} to examine whether the PH
assumption is appropriate.  This, however, can pose a heavy computational
burden
and can be very time-consuming when the data size gets large. While efforts
have
been made in fitting Cox model using distributed computing and therefore
reducing the computing time, such as in \cite{wang2018arXiv}, methods for
checking the PH assumption in these settings have not been developed.

%which has not been tackled
%for big data where the data size exceeds a computer's memory.
%In the context of huge number of covariates, \citet{Mittal2014} developed
%efficient approaches to fit Cox models with regularized covariate coefficients
%for fast variable selection, and \cite{wang2018arXiv} proposed a fast
%divide-and-conquer sparse Cox regression for huge datasets with
%moderate-to-large number of covariates. These studies, however, did not check
%the PH assumption. For this assumption, the test of
%\cite{grambsch1994} is the most popular approach as it is a general framework
%that incorporates many other tests and is available in standard software
%packages \citep[e.g.,][]{Rpkg:survival}. For big data exceeding computers'
%memory, its implementation and alternatives have not been studied.

%We focus on the scenario where survival data arrive
%in streams and inferences need to be updated in an online fashion.
% The desire for online updating methods in the survival setting is not
% uncommon.  For example, flight information, such as delay time until
% take-off or cancellation, is available for more than 114,000 commercial
%flights
% scheduled daily around the world~\citep{aviation2018};
% real estate information, such as time on market until sold, is updated
% continuously for the over 6 million homes
% in the real-estate market~\citep{estate2018}.
In this work, we propose a method to test the PH assumption in the
online updating setting, which does not require storage or access to the
historical data. Our approach is an
 application of the divide-and-conquer and online updating
strategies \citep{lin2011,schifano2016} to the streaming survival data setting.
The data is assumed to arrive sequentially in blocks, % each of
% which is well within the memory limit of the computing facility.
an the test statistic is an appropriately aggregated version of the standard
test statistic of \cite{grambsch1994} computed from each block.
%This method requires minimal storage of a few summaries from historical
blocks.
The statistics can be adapted to be based on data in a moving window of certain
size, which may
be more useful in detecting local deviations from the null hypothesis.
A byproduct of our method is a cumulatively updated estimating equation
(CUEE) estimator for the regression coefficients if the PH
assumption is not rejected.

When the null hypothesis of PH is true, our test statistic
is shown to have the same asymptotic distribution as the standard (full data)
statistic under certain regularity conditions. In simulation studies,
under the null hypothesis, the proposed test holds its size and the
CUEE estimator closely approximates the estimator based on the full data; when
the null hypothesis is not true, the test has comparable or higher power than
the standard statistic based on the full data. For a dataset
that can be loaded into computer memory, our proposed statistic can be computed
in significantly less time than the standard statistic.  Our test can also
successfully be used
within a reasonable amount of time for big data that cannot (easily) be loaded
into memory.
The method is
illustrated by analyzing the survival time of the lymphoma cancer
patients in the Surveillance, Epidemiology, and End Results (SEER) Program.
Interestingly, while the changes in parameters were not captured by using the
standard (full data) test of \cite{grambsch1994}, they were promptly
identified by our online updated version.

The rest of this article is organized as follows. In Section~\ref{sec:cox}, we
review the notation of the Cox model and the test statistic of
\cite{grambsch1994}. In Section~\ref{sec:online}, we propose our online
updating
test statistics for the PH assumption.
We present simulation results in Section~\ref{sec:simu}, and illustrate
the usage of the test with an application to the survival time of patients
with lymphoma cancer from the SEER data in Section~\ref{sec:SEER}.
A discussion concludes in Section~\ref{sec:disc}.
The proposed methods are all implemented in \proglang{R} based on functions
from the \pkg{survival} package \citep{Rpkg:survival}, and the code
can be found via GitHub \citep{xue2018}.

\section{Cox Proportional Hazards Model}\label{sec:cox}
\subsection{Notation and Preliminaries}

For completeness we review the Cox model and tests for the PH assumption.
Let $T_i^*$ be the true event time and $C_i$ be the censoring time for
subject~$i$.
Define $T_i = \min(T_i^*, C_i)$ and $\delta_i = I(T_i^* \leq C_i)$.
Suppose we observe independent copies of
$(\delta_i, T_i, \X_i)$, $i=1,\ldots,n$,
where $\X_i$ is the $p$-dimensional vector of covariates of the $i$th subject.
The Cox model specifies the hazard for individual $i$ as
\begin{equation}\label{eq:defcox}
\lambda_i(t) = \lambda_0(t) \exp\left(\X_i^\top \bbeta\right),
\end{equation}
where $\lambda_0$ is an unspecified non-negative function of time called the
baseline hazard, and $\bbeta$ is a $p$-dimensional coefficient vector in a
compact parameter space. Because the logarithm of the hazard ratio for two
subjects with fixed covariate vectors $\X_i$ and $\X_j$, $(\X_i - \X_j)^\top
\bbeta$, is proportional to the difference in covariate values and is otherwise
constant over time ($\bbeta$), the model is also known as the PH model. It has
been later extended to incorporate time-dependent covariates. For the rest of
the article, we use $\X_i(t)$ to indicate the possibility of covariates being
time-dependent.

\cite{cox1972,cox1975} formulated the partial likelihood approach to
estimate $\bbeta$. For untied failure time data, \cite{fleming1991}
expressed it under the counting process formulation to be
\begin{equation}\label{eq:plklhd}
\mathrm{PL}(\bbeta) = \prod_{i=1}^n \prod_{t\geq 0} \left[
\frac{Y_i(t)\exp\left\{\X_i(t)^\top\bbeta\right\}}
{\sum_{j} Y_j(t)\exp\left\{\X_j(t)^\top\bbeta\right\}}\right]^
{\dd N_i(t) },
\end{equation}
where $Y_i(t) = I(T_i \geq t)$ is the at-risk indicator of the $i$th subject,
$N_i(t)$ is the number of events for subject $i$ at time $t$, and
$\dd N_i(t) = I(T_i\in [t, t + \Delta), \, \delta_i = 1)$, with $\Delta$
sufficiently small such that $\sum_{i=1}^n \dd N_i(t) \leq 1$ for any $t$.
Taking the natural logarithm of \eqref{eq:plklhd} gives the log partial
likelihood in the form of a summation:
\begin{equation}\label{eq:partiallik}
pl(\bbeta) = \sum_{i=1}^n \int_0^\infty \left[Y_i(t)
\exp\left\{\X_i(t)^\top\bbeta\right\}
- \log \sum_{j=1}^n Y_j(t)\exp\left\{\X_j(t)^\top\bbeta\right\} \right]\dd
N_i(t).
\end{equation}
We differentiate $pl(\bbeta)$ with respect to $\bbeta$ to obtain the
$p\times 1$ score vector, $\U(\bbeta)$:
\begin{equation*}
\U(\bbeta) = \sum_{i = 1}^n \int_0^\infty \left\{\X_i(t) -
\bar{\X}(\bbeta, t)\right\}\dd N_i(t),
\end{equation*}
where $\bar{\X}(\bbeta, t)$ is a weighted mean of $\X_i$'s for those
observations still at risk at time $t$ with the weights being their
corresponding
risk scores, $\exp\{\X_i(t)^\top\bbeta\}$. Taking the negative second order
derivative of $pl(\bbeta)$ yields the observed information matrix
$\I_n(\bbeta) = \sum_{i=1}^n \int_0^\infty \V(\bbeta, t) \dd N_i(t) $,
with $\V(\bbeta, t)$ being the weighted variance of $\X$ at time $t$:
\begin{equation*}
\V(\bbeta, t) = \frac{\sum_{i=1}^n
Y_i(t)\exp\{\X_i(t)^\top\bbeta\}\{\X_i(t) - \bar{\X}
(\bbeta, t)\}\{\X_i(t) - \bar{\X}(\bbeta, t)\}^\top }{\sum_i
Y_i(t)\exp\{\X_i(t)^\top\bbeta\}}.
\label{eq:vmatrix}
\end{equation*}
The maximum partial likelihood estimator $\hat{\bbeta}_n$
is obtained as the solution of $\U(\bbeta) = \0$.
The solution $\hat{\bbeta}_n$ is consistent, and asymptotically normal. The
inverse of the observed information, $\I_n (\hat{\bbeta}_n)$, is often
used to approximate the asymptotic variance of $\hat{\bbeta}_n$.

\subsection{Test Statistic for Entire Dataset}\label{ssec:wholeTG}

Following \cite{grambsch1994}, an alternative to PH in
Model~\eqref{eq:defcox} is to allow time-varying coefficients, which can be
characterized by
\begin{equation}\label{eq:h1hazardind}
\beta_j(t)\equiv \beta_j + \theta_j g_j(t),~~j=1,\ldots,p,
\end{equation}
where $g_j(t)$ is a function of time that varies around 0 and $\theta_j$ is
a scalar.
Common choices of $g(t)$ include the Kaplan--Meier~(KM) transformation, which
scales the horizontal axis by the left-continuous version of the KM survival
curve,
the identity function, and the natural logarithm function.
Formulation~\eqref{eq:h1hazardind} is rather general, as
many tests fall within this framework for different choices of $g(t)$
\citep[see, e.g.,][]{xue2017}. Writing \eqref{eq:h1hazardind}
in matrix notation yields
\begin{equation}\label{eq:h1hazard}
\lambda_i(t) = \lambda_0(t) \exp\left[\X_i(t)^\top \{\bbeta +
\G(t)\btheta\}\right], ~~
i = 1,\ldots, n,
\end{equation}
where $\G(t)$ is a $p\times p$ diagonal matrix with the $j$th diagonal element
being $g_j(t)$, and $\btheta = (\theta_1,\ldots, \theta_p)^\top$.
Then the null hypothesis of $\bbeta$ being time-invariant becomes
$H_0: \btheta = \0_{p\times 1}$.

The test of \cite{grambsch1994} is based on Schoenfeld residuals.
Assuming no tied event times and denoting them in increasing order as
$t_1, \ldots, t_d$, where $d$ is the total number of events among the
$n$ observations, the Schoenfeld residuals are defined as
\begin{equation*}
\br_\ell(\bbeta) = \X_{(\ell)} - \bar{\X}(\bbeta, t_\ell),
\end{equation*}
where $\X_{(\ell)}$ is the covariate vector corresponding to the $\ell$th event
time.
In practice,
we use $\hat{\bbeta}_n$ and obtain $\hat{\br}_\ell$ for $\ell = 1,\ldots, d$.
Let $\hat{\V}_{\ell} = \V(\hat{\bbeta}_n, t_\ell)$,
$\G_\ell = \G(t_\ell)$, and
$\bH = \sum_{\ell=1}^d \G_\ell \hat{\V}_\ell \G_\ell - \left(
\sum_{\ell=1}^d \G_\ell \hat{\V}_\ell\right)
\left(\sum_{\ell=1}^d \hat{\V}_\ell\right)^{-1}
\left(\sum_{\ell=1}^d \G_\ell \hat{\V}_\ell\right)^\top.$
\cite{grambsch1994} proposed the statistic
\begin{equation}\label{eq:tg}
T(\G) = \left(\sum_{\ell=1}^d \G_\ell \hat{\br}_\ell\right)^\top \bH^{-1}
\left(\sum_{\ell=1}^d \G_\ell \hat{\br}_\ell\right),
\end{equation}
which, under the null hypothesis, has asymptotic distribution $\chi^2_p$.

For identifiability, $g(t)$ is assumed to vary around 0, so for data analysis
$\G_\ell$, $\ell = 1,\ldots, d$, need to be centered such that
$\sum_{\ell=1}^d\G_\ell=\0_{p\times p}$.
As pointed out by \cite{therneau2000},
$\hat{\V}_\ell$ is rather stable for most datasets, and therefore
$\sum_{\ell=1}^d \G_\ell \hat{\V}_\ell$ is often small. Therefore,
$\bH$ is often replaced by
$\sum_{\ell=1}^d \G_\ell \hat{\V}_\ell \G_\ell$.
The \proglang{cox.zph()} function in the \pkg{survival} package
implements the test in~\eqref{eq:tg} using this same centering technique.
In the sequel, we will assume that all $\G$ matrices are centered
prior to any calculation of the diagnostic statistics.

Tied events are common in practice and there are several
methods to handle ties. We use the approximation of
\cite{efron1977}, which is the default option in the package
\pkg{survival} and returns fairly accurate results
\citep[Section~3.3]{therneau2000}.

\section{Online Updated Test and its Variations}\label{sec:online}

\subsection{Cumulative Version}\label{ssec:cumulative}

Instead of a given, complete dataset, we now consider a scenario in which
survival data become available in blocks. Suppose that for each new arriving
block $k$, we observed $d_k$ events among $n_k$ subjects, for $k=1,\ldots, K$,
where $K$ is some terminal accumulation point of
interest.
With a given $g(t)$ we obtain $d_k$ centered $p\times p$ diagonal matrices
 $\G(t_1),\ldots, \G(t_{d_k})$ such that
$\sum_{\ell=1}^{d_k} \G(t_\ell) = \0_{p\times p}$.
Let $\G_{\ell k}$ and $\hat{\br}_{\ell k}$, $ \ell = 1,\ldots, d_k,$ be the
$k$th block counterpart of previously defined $\G_\ell$  and Schoenfeld
residual
$\hat{\br}_\ell$, respectively.
Without loss of generality, we assume that there is at least
one event in each block so that a Cox model can be fitted,
and each block-wise observed information matrix
$\I_{n_k, k},$ evaluated at some estimate of $\bbeta$, is
invertible.
Let $\V_{\ell k}$ be the weighted variance-covariance matrix of the covariate
matrix at the $\ell$th event time in the $k$th block.
With the approximation that $\hat{\V}_{\ell k} =
\I_{n_k,k}/ d_k$, again where $\I_{n_k,k}$
is evaluated at some estimate of $\bbeta$,
we have $\sum_{\ell = 1}^{d_k} \G_{\ell k}\hat{\V}_{\ell k} = \0_{p\times p}$.
We will discuss the choice of estimate for $\bbeta$ that will be used to
evaluate
$\I_{n_k,k}$, and also $\hat{\br}_{\ell k}$, in Section
\ref{sec:evbeta}.

We denote $\bH_{d_k,k} =
(\sum_{\ell=1}^{d_k} \G_{\ell k} \I_{n_k,k} \G_{\ell k}) / d_k$, and
$\Q_{d_k,k} = \sum_{\ell=1}^{d_k} \G_{\ell k}\hat{\br}_{\ell k}$.
Let $\bH_0 = \0_{p\times p}$, $\bH_{k-1} = \sum_{i=1}^{k-1}\bH_{d_i,i}$,
$\Q_0 = \0_{p\times 1}$, and $\Q_{k-1} = \sum_{i=1}^{k-1}\Q_{d_i, i}$.
Then we have the online updating test statistic given by
\begin{equation}\label{eq.olcum.stat}
T_k(\G) = \Q_k^\top \bH_k^{-1}\Q_k
= (\Q_{k-1} + \Q_{d_k,k})^\top (\bH_{k-1} + \bH_{d_k,k})^{-1}(\Q_{k-1} +
\Q_{d_k,k}).
\end{equation}
At each accumulation point $k$, we need to store
$\bH_{k-1}$ and $\Q_{k-1}$ from previous calculations, and compute
$\bH_{d_k,k}$
and $\Q_{d_k,k}$ for the current block.

\subsection{Window Version}\label{ssec:window}
The cumulative test statistic takes all historical blocks into consideration,
one potential problem of which is that discrepancies from the PH
assumption will accumulate and after a certain time period, the test
will always reject the null hypothesis.
This motivates us to focus on more recent blocks in some applications.
At block $k$, we consider a window of width $w(\geq 1)$, which is tunable,
and
use summary statistics for all blocks in this window to construct the
corresponding test statistic. With $\bH_{d_k,k}$
and $\Q_{d_k,k}$ defined above, we again assume there is at least one event in
each block of data. Denoting $\bH^w_{k} = \sum_{i=k+1-w}^{k} \bH_{d_i,i}$, and
$\Q^w_{k} = \sum_{i=k+1-w}^{k} \Q_{d_i,i}$, the window version online updating
test statistic for nonproportionality based on the most recent $w$ blocks is:
\begin{equation}\label{windowstat}
T^w_k(\G) = (\Q^w_{k})^\top (\bH^w_k)^{-1} \Q^w_k.
\end{equation}
In implementation, we only need to store $\bH_{d_k, k}$ and $\Q_{d_k,k}$ for
all
but the first block in the window,
and compute these summary statistics for
the current block to obtain the aggregated diagnostic statistic. Compared
to the cumulative version statistic, which at each update requires storage of
one $p\times 1$ vector $\Q_k$, one $p\times 1$ vector for an estimate of
$\bbeta$, one $p\times p$ matrix $\bH_k$, and one $p\times p$ variance matrix
of
$\bbeta$, the window version requires storage of these quantities for $w-1$
steps, which is still minimally storage intensive when $p \ll n_k$. 
In addition, as an auxiliary approach that provides an indication approximately
where
along the stream a violation has occurred, $w$ is generally chosen to not be
large. This also makes the storage of these quantities affordable, and the
handling of large blocks possible.

\subsection{Where to Evaluate the Matrices and Residuals}\label{sec:evbeta}

The observed information matrix $\I_{n_k,k}$ and the residuals
$\hat{\br}_{\ell k}$ must be evaluated at a particular choice of $\bbeta$.
A straightforward choice would be $\hat{\bbeta}_{n_k,k}$, the estimate of
$\bbeta$ using the $k$th block of data, $k = 1, 2, \ldots,K$.
It may, however, be more advantageous to use an estimate that utilizes all
relevant historical information.

% Suppose now we have $K$ subsets of data.
Now let us consider the $k$th accumulation point.
The score function for subset $k$ can be obtained as $\U_{n_k,k}(\bbeta)$, and
we
denote the solution to $\U_{n_k,k}(\bbeta) = \0_{p\times 1}$ as
$\hat{\bbeta}_{n_k,k}$.
A Taylor expansion of $-\U_{n_k,k}(\bbeta)$ at $\hat{\bbeta}_{n_k,k}$ is given
by
\begin{equation*}
-\U_{n_k,k}(\bbeta) =
\I_{n_k,k}(\hat{\bbeta}_{n_k,k})(\bbeta - \hat{\bbeta}_{n_k,k}) +
\R_{n_k,k}
\end{equation*}
as $\U_{n_k,k}(\hat{\bbeta}_{n_k,k}) = \0_{p\times 1}$ and $\R_{n_k,k}$ is the
remainder
term. Again, without loss of generality, we assume that there is at least
one event in each block, and each $\I_{n_k, k}$ is invertible.

Denote $\I_{n_k,k}(\hat{\bbeta}_{n_k,k})$ as $\hat{\I}_{n_k,k}$.
Similar to the aggregated estimating equation (AEE) estimator of
\cite{lin2011},
which uses a weighted combination of the subset estimators,
an AEE estimator under the Cox model framework may be given by
\begin{equation}\label{eq.beta.aee}
\hat{\bbeta}_{N} = \left(\sum_{k=1}^K
\hat{\I}_{n_k,k}\right)^{-1}
\sum_{k=1}^K \hat{\I}_{n_k,k}
\hat{\bbeta}_{n_k,k},
\end{equation}
which is the solution to
$\sum_{k=1}^K \hat{\I}_{n_k,k}
(\bbeta - \hat{\bbeta}_{n_k,k}) =\0_{p\times 1}$,
with $N$ being the total number of observations at the final
accumulation point $K$.
\cite{schifano2016} provided the variance estimator for the original
AEE estimator of \cite{lin2011}, and under the Cox model framework it
simplifies to $\hat{\A}_{N} = \left( \sum_{k=1}^K
\hat{\I}_{n_k,k}\right)^{-1}$.

Following \cite{schifano2016}, a
cumulative estimating equation (CEE) estimator for $\bbeta$ at
accumulation point $k$ under the Cox model framework is
\begin{equation}\label{eq.betak.cee}
\hat{\bbeta}_k = \left(\hat{\I}_{k-1} +
\hat{\I}_{n_k,k}\right)^{-1}
\left(\hat{\I}_{k-1}\hat{\bbeta}_{k-1} +
\hat{\I}_{n_k,k}\hat{\bbeta}_{n_k,k}\right)
\end{equation}
for $k = 1,2,\ldots$, where $\hat{\bbeta}_0 = \0_{p\times 1}$,
$\hat{\I}_0 = \0_{p\times p}$, and $\hat{\I}_k =
\sum_{i = 1}^k \hat{\I}_{n_i,i} = \hat{\I}_{k-1}
+ \hat{\I}_{n_k,k}$.
The variance estimator at the $k$th update simplifies to
$\hat{\A}_k = \left(\hat{\I}_{k-1} +
\hat{\I}_{n_k,k}\right)^{-1}.$
Note that for terminal $k=K$, the AEE estimators and CEE estimators coincide.

%As pointed out by \cite{schifano2016}, the CEE estimators are not equivalent
to
%the estimating equation (EE) estimators (based on the entire sample) in
%finite sample sizes.
Similar to \cite{schifano2016}, we propose a CUEE estimator %under the EE
framework
to better approximate the maximum partial likelihood estimator (based on the
entire sample) with less bias.
Take the Taylor expansion of $-\U_{n_k,k}(\bbeta)$ around $\check{\bbeta}
_{n_k,k}$, which will be defined later. We have
\begin{equation*}
-\U_{n_k,k}(\bbeta) = -\U_{n_k,k}(\check{\bbeta}_{n_k,k}) +
\I_{n_k,k}(\check{\bbeta}_{n_k,k}) (\bbeta -
\check{\bbeta}_{n_k,k})
 + \check{\R}_{n_k,k},
\end{equation*}
where $\check{\R}_{n_k,k}$ is the remainder term. Again for simplicity, we
denote $\I_{n_k,k}(\check{\bbeta}_{n_k,k})$ as
$\check{\I}_{n_k,k}$,
and $\U(\check{\bbeta}_{n_k,k})$ as $\check{\U}_{n_k,k}$.
We now ignore the remainder term and sum the first order expansions for
blocks $1,\ldots, K$, and set it equal to $\0_{p\times 1}$:
\begin{equation}\label{eq:aee}
\sum_{k=1}^K -\check{\U}_{n_k,k} + \sum_{k=1}^K
\check{\I}
_{n_k,k}\left(\bbeta - \check{\bbeta}_{n_k,k}\right)=\0_{p\times 1}.
\end{equation}
Then we have the solution to \eqref{eq:aee}:
$\tilde{\bbeta}_K = \left(\sum_{k=1}^K \check{\I}_{n_k,k}
\right)^{-1} \left(\sum_{k=1}^K
\check{\I}_{n_k,k}\check{\bbeta}_{n_k,k}
 + \sum_{k=1}^K \check{\U}_{n_k,k}
\right)$.
The choice of $\check{\bbeta}_{n_k,k}$
is subjective. At accumulation point $k$,
it is possible to utilize information at the previous accumulation
point to define $\check{\bbeta}_{n_k,k}$.
One candidate intermediary estimator is
\begin{equation}\label{eq:betacheck}
\check{\bbeta}_{n_k,k} = \left(\check{\I}_{k-1} +
\hat{\I}_{n_k,k}\right)^{-1}
\left(\sum_{i=1}^{k-1}\check{\I}_{n_i,i} \check{\bbeta}_{n_i,i}
 + \hat{\I}_{n_k,k}\hat{\bbeta}_{n_k,k} \right)
\end{equation}
for $k = 1,2,\ldots$, $\check{\I}_0 = \0_{p\times p}$,
 $\check{\bbeta}_{n_0,0}
= \0_{p\times 1}$, and $\check{\I}_k = \sum_{i=1}^k
 \check{\I}_{n_i,i}$.
Estimator \eqref{eq:betacheck} is the weighted combination of
 the previous intermediary estimators
$\check{\bbeta}_{n_i,i}, i = 1,\ldots, k-1$ and the current subset estimator
$\hat{\bbeta}_{n_k,k}$. It results as the solution to the estimating equation
$\sum_{i=1}^{k-1}\check{\I}_{n_i,i} \left(\bbeta -
\check{\bbeta}_{n_i,i}\right) + \hat{\I}
_{n_k,k} \left(\bbeta - \hat{\bbeta}_{n_k,k}\right) = \0_{p\times 1}$, with
 $\hat{\I}_{n_k,k} \left(\bbeta - \hat{\bbeta}_{n_k,k}\right)$
being the bias correction term since $-\sum_{i=1}^{k-1}\check{\U}_{n_i,i}$ has
been omitted.

With $\check{\bbeta}_{n_k,k}$ given in \eqref{eq:betacheck}, the CUEE
estimator $\tilde{\bbeta}_k$ for the Cox model is
\begin{equation*}
\tilde{\bbeta}_k = \left(\check{\I}_{k-1} +
\check{\I}_{n_k,k}\right)^{-1}
\left(\s_{k-1} + \check{\I}_{n_k,k} \check{\bbeta}_{n_k,k} + \bxi_{k-1}
+ \check{\U}_{n_k,k}\right),
\end{equation*}
with $\s_k = \sum_{i=1}^k \check{\I}_{n_i,i}\check{\bbeta}_{n_i,i} =
\check{\I}_{n_k,k}\check{\bbeta}_{n_k,k} + s_{k-1}$ and $\bxi_k =
\sum_{i=1}^k \check{\U}_{n_i,i} = \check{\U}_{n_k,k}
+ \bxi_{k-1}$, where $\s_0 = \bxi_0 = \0_{p\times 1}$,
 and $k = 1,2, \ldots$.
For the variance of $\tilde{\bbeta}_k$, as
$\0_{p\times 1} = -\hat{\U}_{n_k,k} \approx
 - \check{\U}_{n_k,k} +
\hat{\I}_{n_k,k}
\left(\hat{\bbeta}_{n_k,k} - \check{\bbeta}_{n_k,k}\right)$,
we have
$\check{\I}_{n_k,k}\check{\bbeta}_{n_k,k} +
\check{\U}_{n_k,k}
\approx \check{\I}_{n_k,k}\hat{\bbeta}_{n_k,k}$.
The estimated variance of $\tilde\bbeta_k$ is online updated by, in simplified
form,
\begin{equation*}
\widetilde{\mathrm{Var}}(\tilde\bbeta_k) =
\left(\check{\I}_{k-1} + \check{\I}_{n_k,k}\right)^{-1}
\left(\sum_{i=1}^k
\check{\I}_{n_k,k}\hat{\I}_{n_k,k}^{-1}
\check{\I}_{n_k,k}
^\top
\right) \left\{\left(\check{\I}_{k-1} +
\check{\I}_{n_k,k}\right)^{-1}\right\}^\top.
\end{equation*}

Thus, for the cumulative version statistic, the matrices and Schoenfeld
residuals
are evaluated at $\tilde{\bbeta}_k$, the CUEE estimator, in our implementation.
For the window
version statistic, the matrices and Schoenfeld residuals are evaluated
at the CEE estimator, as with a limited window size, there is
little room for the bias of the CEE estimator to accumulate, and the
difference between the CUEE estimator and the CEE estimator within a window is
negligible for
small $w$.
Note that when $w=1$, both estimators are the same, and are equal to the
parameter estimate for
the current block, $\hat{\bbeta}_{n_k,k}$.

\subsection{Asymptotic Results}

We now provide the asymptotic distribution of the test statistic $T_k(\G)$
given in Equation~\eqref{eq.olcum.stat}. For ease of
presentation, we assume that all subsets of data are of equal size $n$, i.e.,
$n_k=n$.

\begin{thm}\label{thm1}
Under conditions C1-C5 in Web Appendix~A, as
$n\rightarrow\infty$, if $K=O(n^\gamma)$ with
$0<\gamma<\min\{1-2\alpha, 4\alpha-1\}$, then for any $k\le K$, the test
statistic satisfies that
\begin{equation*}
    T_k(\G) \rightarrow \chi^2_p,
\end{equation*}
in distribution when all data blocks follow the PH model with the same
covariate parameters.
\end{thm}

The proof is provided in Web
Appendix~A.
The asymptotic
distribution is valid for any stage of the updating process if each
subset is not very small and the null hypothesis is true. This means
that the type one error rate is always
well maintained.
As more data accumulate along the updating procedure,
the test statistic gains more power. If the $n_k$'s are different, the
asymptotic result is still valid under some mild condition, for
example, $\max_k n_k / \min_k n_k = O(1)$. Note that the window version
statistic $T_k^w(\G)$ is essentially the cumulative version statistic evaluated
at the CEE with different starting blocks. Therefore, the asymptotic
distribution is also valid for the window version statistic. In the special
case
of $w=1$, the proposed
statistic reduces to the original $T(\G)$ on the most recent block, which has
been shown to be $\chi^2_p$ by \cite{grambsch1994}.

\section{Simulation Studies}\label{sec:simu}

Simulation studies were carried out to evaluate the empirical sizes and powers
of both $T_k(\G)$ and $T_k^w(\G)$.
When data were generated under the PH assumption,
we also compared the empirical distribution of $T_k(\G)$ with that of the
standard statistic computed using all data
up to selective accumulation points $k$, denoted by $T_{1:k}(\G)$.
While we look at the end of each stream to decide whether the entire stream
of data satisfies the PH assumption or not, we also examine the
results at each accumulation point to verify the performance of the proposed
test statistics. Simulations have also been conducted to assess the savings
in computing time and reduction in memory usage for the proposed statistics
with
big survival data.
See Web Appendices~B.1
and~B.2.

\subsection{Size}\label{ssec:null}

Event times were generated from Model~\eqref{eq:defcox} with three covariates
$x_{ki[1]}\overset{\textrm{i.i.d.}}{\sim} N(0,1)$,  $x_{ki[2]}
\overset{\textrm{i.i.d.}}{\sim} \textrm{Bernoulli}(0.5)$,
$x_{ki[3]}\overset{\textrm{i.i.d.}}{\sim}\textrm{Bernoulli}(0.1)$ for $i =
1,\ldots,
n_k$, making a $n_k\times 3$ covariate matrix. We set a vector of
parameters
$\bbeta_0 = (0.67,-0.26, 0.36)^{\top}$,
and baseline hazard $\lambda_0(t) = 0.018$.
Censoring times were generated independently from a mixture distribution:
$\varepsilon \langle 60 \rangle +
(1-\varepsilon)\mathscr{U}(0,60)$, where $\langle 60 \rangle$ represents
a point mass at 60, and $\mathscr{U}(0,60)$ denotes the uniform distribution
over (0, 60).
Setting $\varepsilon=0.9$ gives approximately 40\%
censoring rate, and $\varepsilon=0.1$ gives approximately 60\% censoring rate.
For each censoring level, we generated $1,000$ independent streams of survival
datasets, each of which had $N = 200,000$ observations in $K = 100$ blocks
with $n_k = 2,000$.

\begin{figure}[tbp]
 \centering
\includegraphics[width = \linewidth]{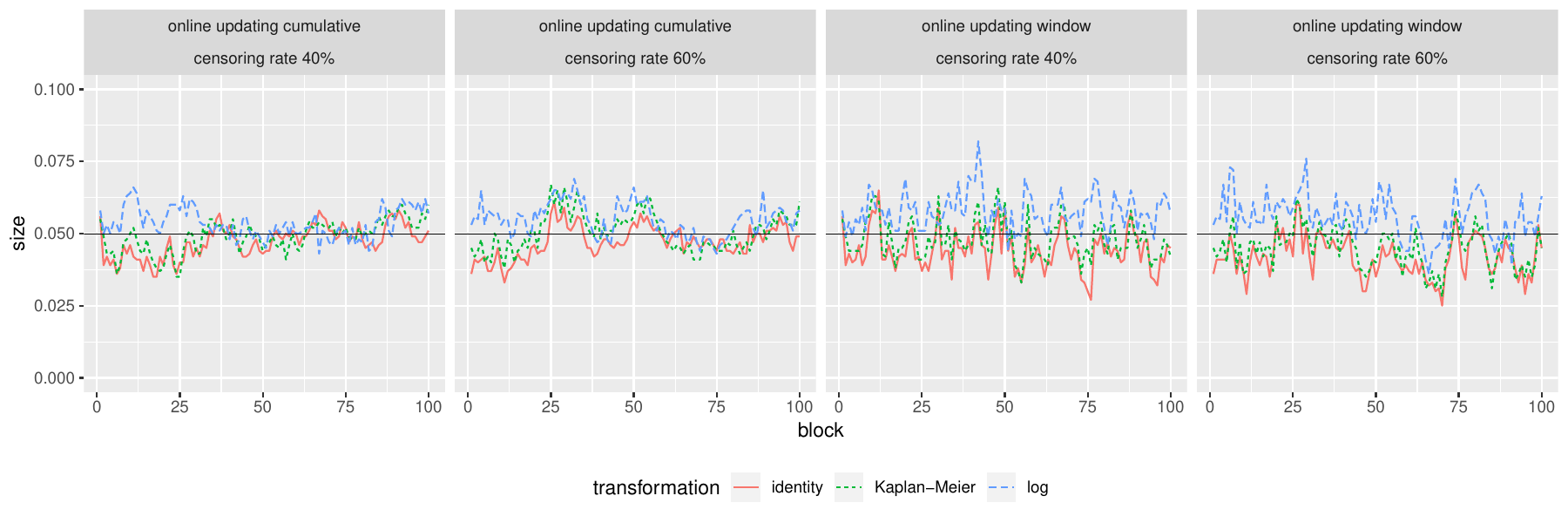}
\caption{Empirical size (proportion of statistic values greater than
$\chi^2_{3, 0.95}$) calculated at each update using the identity,
KM, and log transformations under the null hypothesis. This figure appears in
color in the electronic version of this article.}
\label{fig:nullplots}
\end{figure}

Three choices of $g(t)$ were considered, the identity, KM, and log
transformations, in the calculation of the test statistics. For each choice,
we calculated both $T_k(\G)$ and  $T_k^w(\G)$ with $w= 5$
upon arrival of each block of simulated data.
Figure~\ref{fig:nullplots} summarizes empirical sizes of the
test
with nominal level 0.05 at each accumulation point $k = 1, \ldots, 100$
for the two versions of the tests under two censoring levels.
The empirical sizes for the three choices of $g(t)$ fluctuate
closely around the nominal level 0.05 in all the scenarios. The log
transformation, however, results in a slightly larger size,
and its usage should therefore be treated with caution.

\begin{figure}[tbp]
\centering
\includegraphics[width = \linewidth]{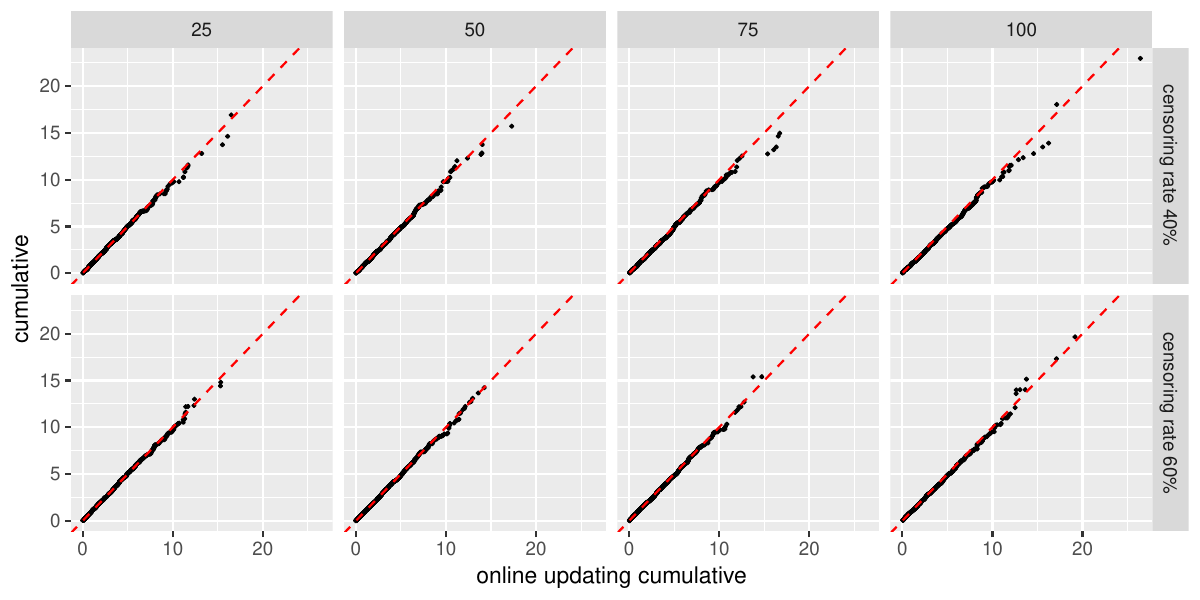}
\caption{Empirical quantile-quantile plots of the online updating
cumulative statistics $T_k(G)$ ($x$-axis)
and $T_{1:k}(G)$ obtained using cumulative data ($y$-axis) with censoring rate
40\% and 60\%, taken at block $k \in \{25, 50, 75, 100\}$,
both calculated using the KM transformation
on event times. This figure appears in color in the electronic version of this
article.}
\label{fig:qqplots}
\end{figure}

To compare the empirical distribution of $T_k(\G)$ and the standard statistic
$T_{1:k}(\G)$,
we additionally computed $T_{1:k}(\G)$ at blocks $k \in \{25, 50, 75, 100\}$
based on cumulative data up to those blocks.
Figure~\ref{fig:qqplots} presents the quantile-quantile plots of the two
statistics obtained with $g(t)$ being the KM transformation.
The points line up closely on the 45 degree line, confirming that the
online updating cumulative statistics $T_k(\G)$ follow the same asymptotic
$\chi^2_p$ distribution under the null hypothesis as $T_{1:k}(\G)$.

Additional simulation results on the sizes for scenarios where $p \in
\{10, 20\}$ and where covariate coefficients are piecewise constant with
respect to time (and accommodated in the PH model
by including additional covariates to handle the pieces separately)
are reported in Web Appendices~B.3
and~B.4. In both cases, the size was well-maintained.

\subsection{Power}\label{ssec:power}

Continuing with the simulation setting from Section \ref{ssec:null}, two
scenarios where the PH assumption is violated were considered to assess the
power of the proposed tests.

\begin{figure}[tbp]
 \centering
\includegraphics[width = \linewidth]{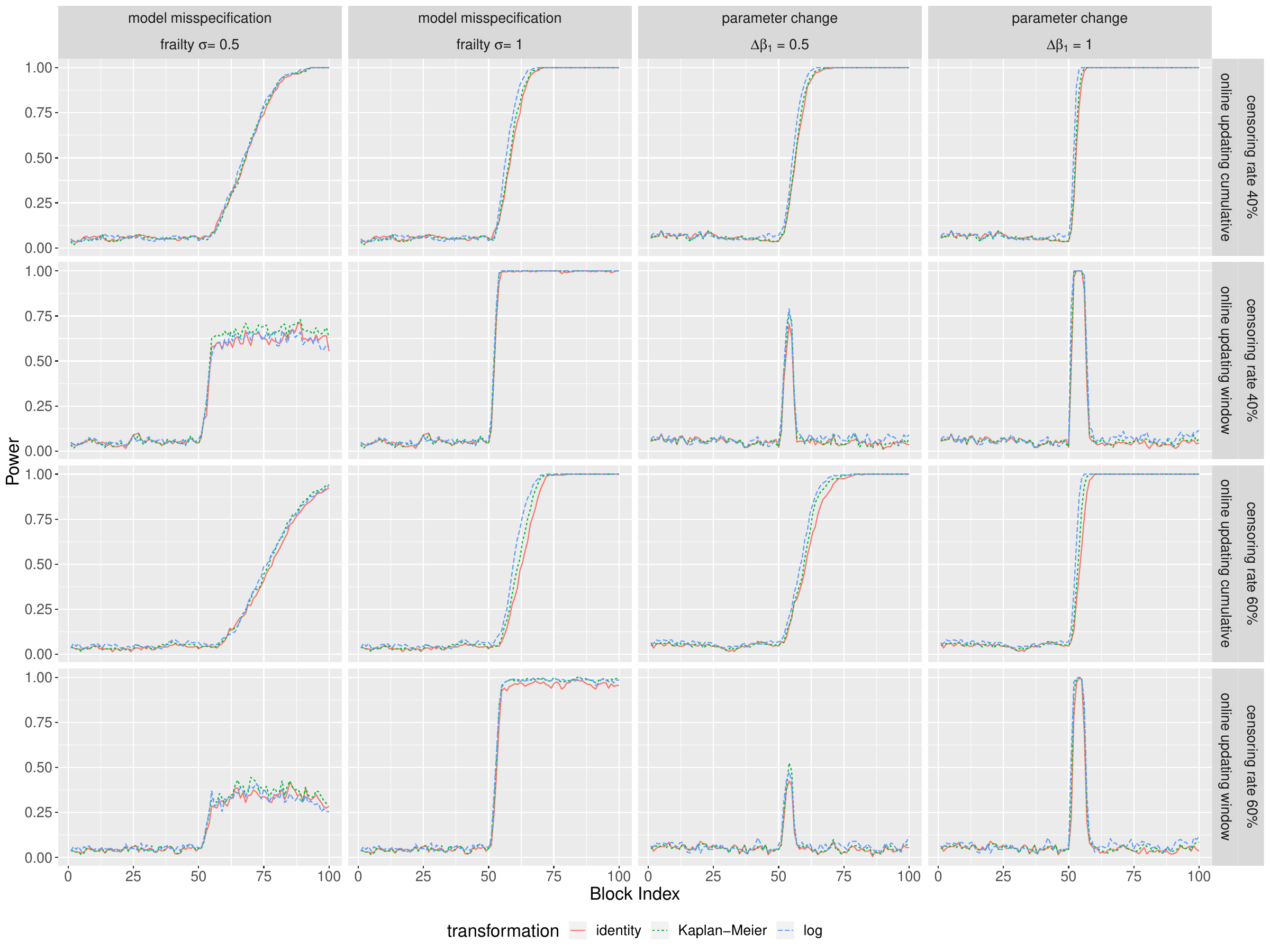}
\caption{Empirical power (proportion of statistic values greater than
$\chi^2_{3, 0.95}$) for the online updating cumulative and window tests,
calculated at each update using the identity,
KM, and log transformations under the alternative hypotheses
 of model misspecification (left) and parameter change (right) under
censoring rate $40\%$ (top) and $60\%$ (bottom). This figure appears in color
in the electronic version of this article.}
\label{fig:alternativeplots}
\end{figure}

The first scenario breaks the PH assumption by a multiplicative
frailty in the hazard function. Starting from the 51st block in each stream,
the hazard function, instead of being~\eqref{eq:defcox}, becomes
$\lambda_i(t)  = \lambda_0(t)\exp(X_i^\top \bbeta + \epsilon_i)$,
where a normal frailty $\epsilon_i \sim N(0,\sigma^2)$ is introduced.
Two levels of $\sigma$ were considered, 0.5 and~1.
Figure~\ref{fig:alternativeplots} shows the empirical rejection rates of the
tests at level 0.05 from 1,000 replicates against accumulation point $k$.
The tests have higher power under lower censoring rate or higher frailty
standard
deviation. At a given censoring rate and frailty standard deviation,
$T_k^w(\G)$
picks up the change more rapidly than~$T_k(\G)$
because it discards information from older blocks for which the
PH assumption holds; the power remains at a certain level
(less than~1) after all the blocks in the window contain data generated from
the frailty model. While $T_k(\G)$ responds to the change more slowly,
as the proportion of blocks with data generated from
the frailty model increases, the power approaches~1 eventually.
In all settings, tests based on the log and KM
transformations seem to have higher power than that based on the identity
transformation.

The second scenario breaks the PH assumption by
a change in one of the covariate effects. Specifically, we considered
an increase of 0.5 or~1 in $\beta_1$, the coefficient for the first
covariate
in data generation,  starting from the 51st block.
The empirical rejection rates of the tests with level 0.05 from 1,000
replicates are presented in Figure~\ref{fig:alternativeplots}.
Both versions of the tests have higher power when the censoring rate is
lower or the change in $\beta_1$ is larger.
At a given censoring rate and change in $\beta_1$, $T_k^w(\G)$ only has
power to detect the change near the 51st block, where the blocks in the
window contain data from two models.
The cumulative version, $T_k(\G)$, picks up the change after the 51st block and
the power increases quickly to~1.

A more comprehensive simulation study was conducted to compare the power of
$T_K(\G)$ and the full data test statistic $T(\G)=T_{1:K}(\G)$ at the end of
each data stream, and the results are presented in Web
Appendix~B.5. When there is a
model change, the power of $T_K(\G)$ is comparable to the power of $T(\G)$;
when there is a change in covariate effect, $T_K(\G)$ has significantly higher
power than $T(\G)$.

\section{Survival Analysis of SEER Lymphoma Patients}
\label{sec:SEER}

We consider analyzing the survival time of the lymphoma patients in the
SEER program with the proposed methods. Among the 131,960 patients diagnosed
with lymphoma between 1973 to 2007, 47,009 experienced an event within 60
months due to lymphoma, resulting in a censoring rate of 64.4\%.
The risk factors
considered in our analysis were \proglang{Age} (centered and scaled),
gender indicator (\proglang{Female}),
African-American indicator (\proglang{Black}).
There were 60,432 females, and 9,199 African-Americans.
%While the dataset is large, the analysis of the data as a single dataset is
%still possible with
%reasonable computing resources.
We wish to compare the performance of the
standard statistic $T(\G)$ from Equation~\eqref{eq:tg} with $T_k(\G)$ under a
setting in which the PH assumption is judged to be satisfied based on the
standard $T(\G)$ test.
For online updating, the patients in the data were ordered by time of
diagnosis, and partitioned by quarter of a year into 140 blocks.
The average sample size per block was 943, but the block sizes and censoring
rates increased over time; see Web Figure~S4.

As a starting point, an initial model that included the three risk factors
was fitted, and $T(\G)$ based on the full data as in
Equation~\eqref{eq:tg} was calculated to be 83.38, which indicated that the
model does not satisfy the PH assumption. The online
updating cumulative statistic $T_k(\G)$ was calculated to be 95.60.
Due to the relatively high censoring rate, the KM transformation was chosen in
calculation of the diagnostic statistics
as it is more robust in such a scenario \citep[e.g.,][]{xue2017}.
Diagnosis with function \proglang{plot.cox.zph()} in the \pkg{survival}
package revealed that all the parameters are likely to be time-dependent;
see Web Figure~S5.

Techniques in \cite{therneau2017} were used to allow the parameters to be
piecewise-constant over time. Two cut-offs were chosen at 2 and 30 months
based on the time-variation pattern of $\hat{\bbeta}(t)$
obtained from the naive model. A
factor variable \proglang{tgroup} is defined to indicate on which
intervals the corresponding observation contributes to estimation of $\bbeta$.
For example, a subject with survival time 25 and event 1 will now be
represented separately on two intervals:
one with time interval $(0, 2]$, with event 0 and
$\proglang{tgroup}=1$, and the other with time interval $(2,25]$, with
event 1 and $\proglang{tgroup}=2$.
The interaction of \proglang{Age}, \proglang{Female}
and \proglang{Black} with the generated \proglang{tgroup} as strata gives the
model more flexibility to fit to the data.
The new model resulted in $T(\G)=T_{1:140}(\G)= 5.75$ on 9 degrees of freedom
with a $p$-value of 0.77, which indicates that the PH
assumption for the revised model is appropriate based on the full data.
Web Figure~S6 presents time-variation plot of
parameters for the revised model.

\begin{figure}[tbp]
 \centering
\includegraphics[width = \linewidth]{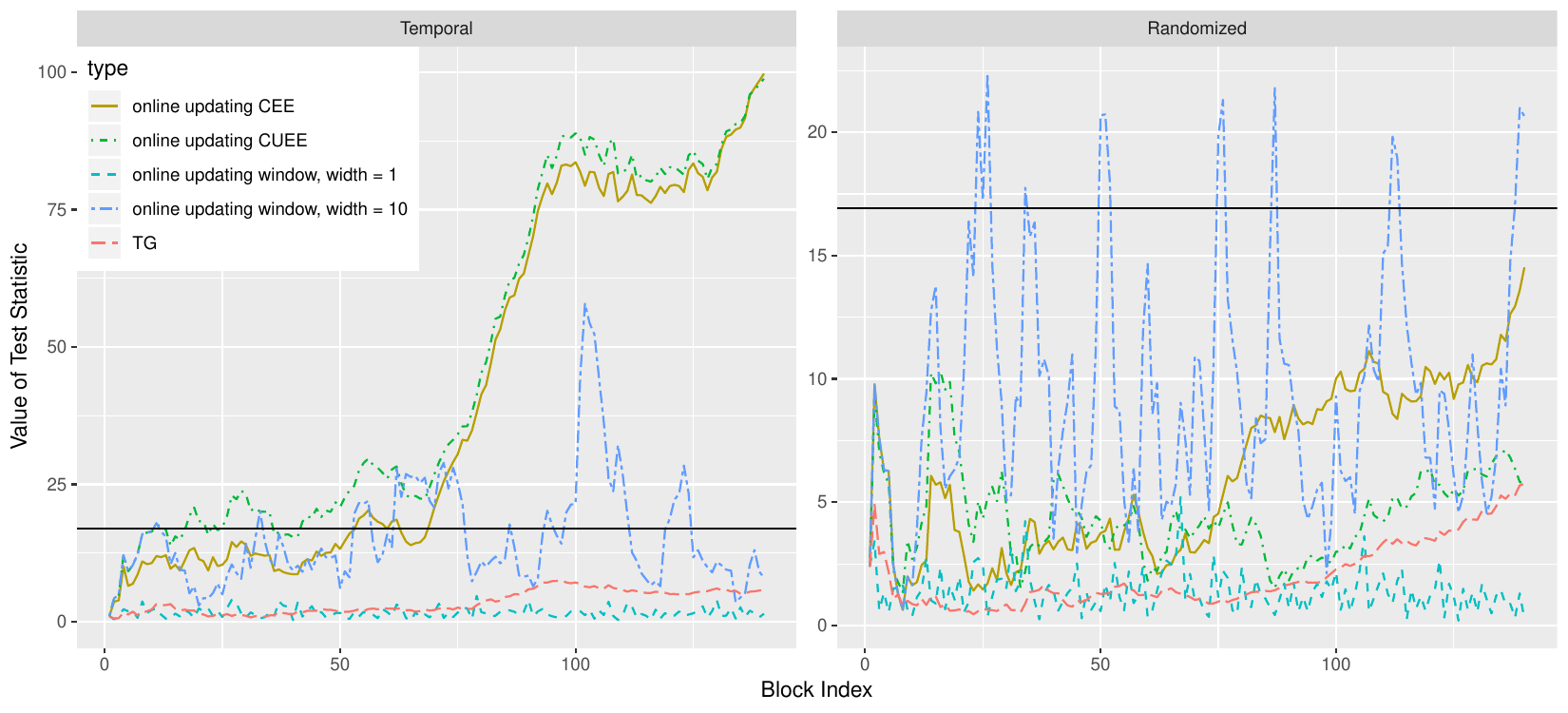}
\caption{Test statistics for the PH assumption for lymphoma
  data, using temporally ordered (left) and randomly ordered (right)
datasets. This figure appears in color in the electronic version of this
article.}
\label{fig:diagstats}
\end{figure}

To evaluate the performance of the online updating parameter estimates
and test statistics under the revised model,
at each block $k, ~k = 1,\ldots, 140$, we calculated the parameter estimates,
$T_k(\G)$, $T_k^w(\G)$, and also $T_{1:k}(\G)$ based on the
single large dataset consisting of all cumulative data up to block~$k$.
Two versions of $T_k(\G)$ were obtained, one using
the CEE estimator $\hat{\bbeta}_k$ and the other using the CUEE estimator
$\tilde{\bbeta}_k$.  For $T_k^w(\G)$, the CEE estimator $\hat{\bbeta}_k$ was
used as discussed previously, and two widths $w=1$ and $w=10$ were considered.
The trajectories of different versions of the test statistics were plotted in
the left panel of Figure~\ref{fig:diagstats}. While the PH assumption seemed to
be satisfied within each individual block ($w=1$), as well as in
cumulative data up to each accumulation point, both online updating cumulative
statistics $T_k(\G)$ resulted in a rejection of the null hypothesis,
and $T_k^w(\G)$ when $w=10$ also resulted in a few rejections along the stream.

\begin{figure}[t]
    \centering
    \includegraphics[width = \linewidth]{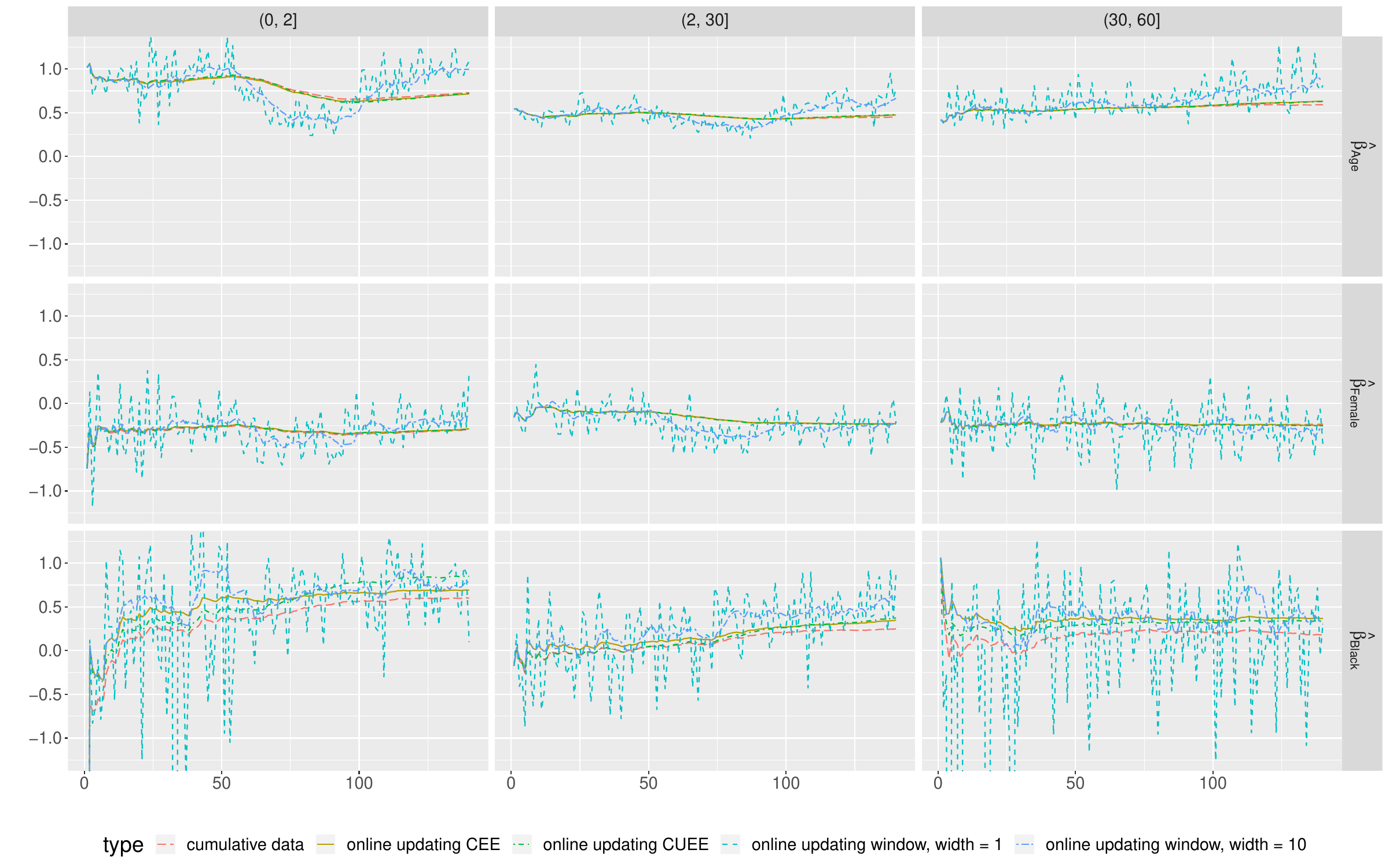}
\caption{Parameter estimates for \textsf{Age}, \textsf{Female} and
\textsf{Black} on intervals $(0,2]$, $(2, 30]$ and $(30, 60]$, given by
different estimating schemes on the temporally ordered lymphoma dataset,
plotted against block indices. The decreasing and then increasing trend in the
first piece of $\hat{\beta}_{Age}$ is clear. This figure appears in color in
the electronic version of this article.}
    \label{fig:betahat}
\end{figure}

The trajectories of three parameter estimates $\hat{\beta}_{\textsf{Age}}$,
$\hat{\beta}_{\textsf{Female}}$, and $\hat{\beta}_{\textsf{Black}}$ on
the three time intervals $(0,2]$, $(2, 30]$ and $(30,60]$
(obtained from the covariate interactions with \proglang{tgroup})
 were plotted with respect to block indices to investigate this
apparent discrepancy; see Figure~\ref{fig:betahat}. Apparently,
$\hat{\beta}_{\textsf{Age}}$ on $(0,2]$ remained relatively stable for blocks 1
to 50, but started to first decrease and later increase.
This change was captured by both $T_k^w(\G)$ and $T_k(\G)$, but not by
$T_{1:k}(\G)$. This is explained by the
fact that $T_{1:k}(\G)$ is based on a single estimator of $\bbeta$, while in
the online updating statistics, each block has its own estimate of $\bbeta$.
The temporal changes that are observed in the CUEE estimate of $\bbeta$
get canceled in the calculation based on the full cumulative data.

\begin{figure}[t]
    \centering
    \includegraphics[width = \linewidth]{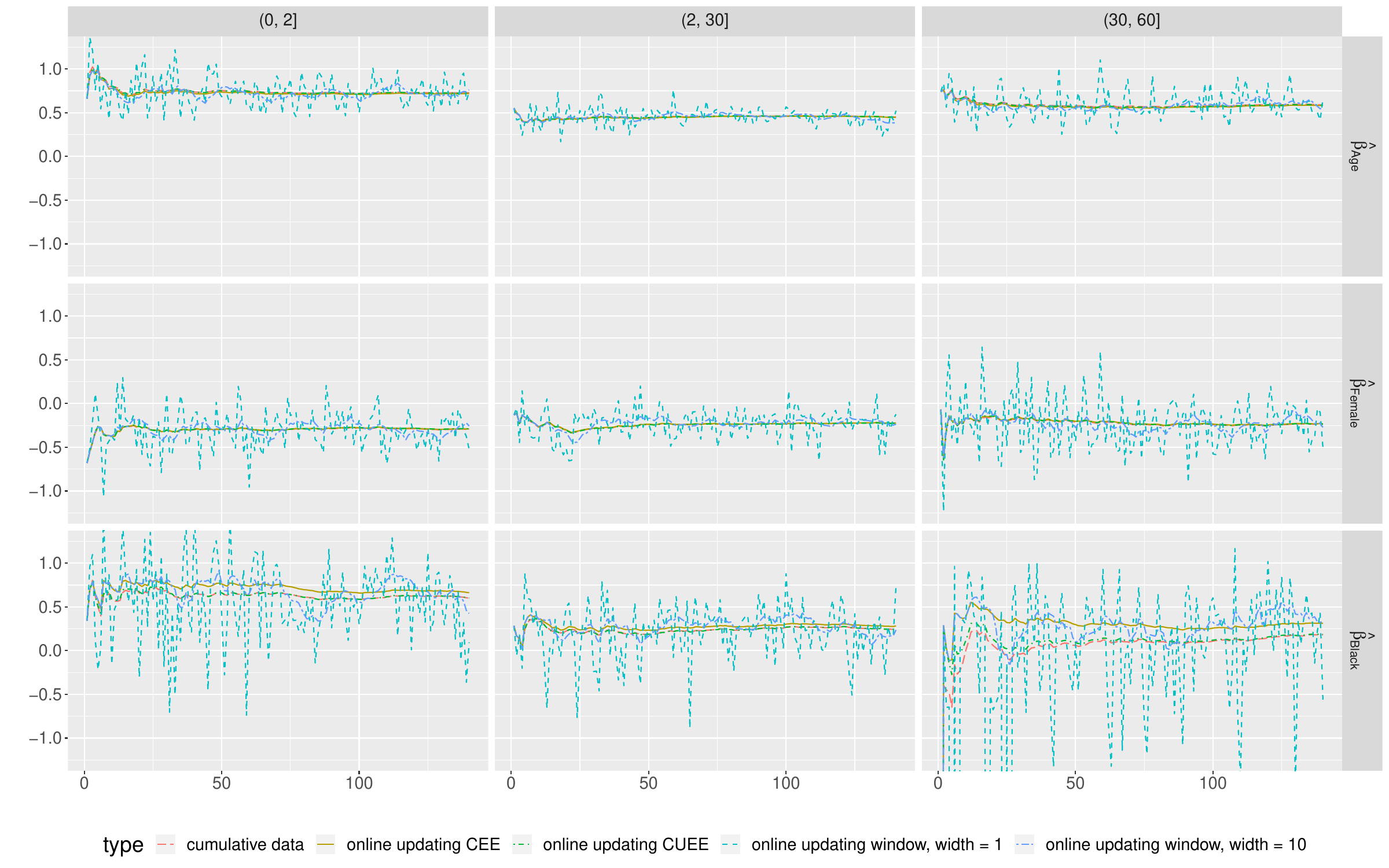}
    \caption{Parameter estimates for \textsf{Age}, \textsf{Female} and
\textsf{Black} on intervals $(0,2]$, $(2, 30]$ and $(30, 60]$, given by
different estimating schemes on the randomly ordered lymphoma dataset,
plotted against block indices. While the blockwise parameter estimates
(window version with $w=1$) are still volatile, there is no significant
increasing or decreasing trend from older blocks to newer blocks. This figure
appears in color in the electronic version of this article.}
    \label{fig:betahatrandom}
\end{figure}

To confirm that the temporal change in parameter contributed to the highly
significant online updating test statistics, we randomly permuted the order of
the observations in the original dataset 1,000 times using the same block
size as the original data.
For each permutation, we applied the same techniques and cut-offs to allow for
piecewise constant parameters over time as before.
The histogram of the 1,000 CUEE-based $T_k(\G)$ is included in Web
Figure~S7. The empirical $p$-value based on these 1,000
permutations is 0.016, indicating that the particular order of blocks in
the original temporally ordered data is indeed contributing to
non-proportionality.
Figure~\ref{fig:betahatrandom} presents the same diagnostic plots as
Figure~\ref{fig:betahat} except that they are for one random permutation.
While the final cumulative data parameter estimates remain the same, the
trajectories are much flatter, with no obvious temporal trend over blocks.
The diagnostic statistics were also obtained under this random permutation,
and plotted in the right panel of Figure~\ref{fig:diagstats}. Each block again
satisfies the PH assumption, and the performance of the
online updating cumulative statistic based on CUEE is very close to
$T(\G)$ computed on the entire dataset. The online updating window version
($w=10$),
however, still identified a few neighborhoods where the variation is large,
and this behavior persists across different choices of window size.

\section{Discussion}\label{sec:disc}

We developed online updating test statistics for the PH
assumption of the Cox model for streams of survival data.
The test statistics were inspired by the divide and conquer approach
\citep{lin2011} and the online updating approach for estimation and inference
of regression parameters for estimating equations \citep{schifano2016}.
We proposed two versions of test statistics, $T_k(\G)$ using cumulative
information from all historical data, and $T_k^w(\G)$ using
information only from more recent data. Both statistics have an
asymptotic~$\chi^2_p$ distribution under the null hypothesis.
In our simulation studies,
the power of $T_k(\G)$ is comparable to or higher than the power of the
standard test $T(\G)$ on the entire dataset,
for scenarios of a model change or parameter change,
respectively. In addition, when $T(\G)$ fails to detect violation of the
null hypothesis on the whole dataset, $T_k(\G)$ may still identify the
violation with high power. This was observed in our application to the SEER
data, and also echoes the findings in \cite{Battey:wf}.
This also suggests that, even when the dataset is not huge,
it might be desirable to partition the data and examine the partitions
for possibly masked violations of the null hypothesis. At the final block,
the cumulative version test statistic will help us decide if the PH assumption
has been satisfied. The window version, however, can be run at the same
time, as it is sensitive to heterogeneity among a few blocks.

As with previous online updating approaches, $T_k(\G)$ and $T_k^w(\G)$ are
computationally fast, and minimally storage intensive. As shown in the
supporting information, the methods are also capable
of handling large datasets of a few gigabyte's size, and can return the
estimation
and diagnostic results within reasonable time limit.
%When the dataset is too large to be loaded into memory, our approach can
%still be performed within reasonable time limit.
Compared to parallel computing for such datasets,
the proposed approach reduces time needed for communication between
nodes, and allows for bias correction of the parameter estimates.

A few issues beyond the scope of this paper are worth further investigation.
The size of blocks should be chosen following general
guidelines \citep[e.g.,][]{schoenfeld1983} so that the covariate effects can be
sufficiently identified, and that the information matrices exist and are
invertible. In practice, with a data stream, we can always choose to let the
data accumulate until a certain number of events
are observed. Then these observations can be grouped into one block,
which can produce stable and valid results for test purposes. For $T_k^w(\G)$,
the choice of $w$ may affect the test results and
local parameter estimates. Possible influential factors include the size of
data chunks, the censoring rate within each chunk, among others.
Additionally, as we are more interested in local or current
goodness-of-fit when using the window version, $w$ should generally be small.
Also, as illustrated in Figure~\ref{fig:alternativeplots}, $T_k^w(\G)$ can
behave differently under different violations of the PH assumption,
therefore, prior knowledge on what types of changes are likely to occur, if
available, may also be taken into consideration. As we are more
concerned with deciding whether the entire stream satisfies the PH
assumption, this window version should be treated as of auxiliary
purpose.
Also, the test statistics and parameter estimates perform well when $p$ is
small to moderate. When $p$ is high or ultra-high, singularity issues could
arise, and appropriate penalization methods should
be considered~\citep[e.g.][]{fan2002variable,zou2008,Mittal2014}.

Finally, in this work we are only concerned with making a final decision
regarding the PH assumption at the end of a data stream.
There are scenarios, however, under which we may wish to make decisions
alongside the data stream as the updating process progresses. This brings up
the issue of multiple hypothesis testing. Hypothesis testing in the online
updating framework is an interesting topic, and has been explored
recently in \cite{webb2016} and \cite{javanmard2018}, and also in the
statistical process control framework in, e.g., \cite{lee2010,lee2012}.
Appropriate adjustment procedures in the online updating PH test
context are areas devoted for future research.

\end{document}